# High average power ultrafast laser technologies for driving future advanced accelerators

Leily Kiani[1], Tong Zhou[2], Seung-Whan Bahk[3], Jake Bromage[3], David Bruhwiler[4], E. Michael Campbell[3], Zenghu Chang[4], Enam Chowdhury[5], Michael Downer[6], Qiang Du[2], Eric Esarey[2], Almantas Galvanauskas[7], Thomas Galvin[1], Constantin Häfner[8], Dieter Hoffmann[8], Chan Joshi[9], Manoj Kanskar[10], Wei Lu[11], Carmen Menoni[12], Michael Messerly[1], Sergey B. Mirov[13], Mark Palmer[14], Igor Pogorelsky[14], Mikhail Polyanskiy[14], Erik Power[3], Brendan Reagan[1], Jorge Rocca[12], Joshua Rothenberg[15], Bruno E. Schmidt[16], Emily Sistrunk[1], Thomas Spinka[1], Sergei Tochitsky[9], Navid Vafaei-Najafabadi[17], Jeroen van Tilborg[2], Russell Wilcox[2], Jonathan Zuegel[3], Cameron Geddes[2]

[1]Lawrence Livermore National Laboratory, Livermore, CA, USA
[2]Lawrence Berkeley National Laboratory, Berkeley, CA, USA
[3]University of Rochester, Laboratory for Laser Energetics, Rochester, NY, USA
[4]University of Central Florida, Orlando, FL, USA
[5]Ohio State University, Columbus, OH, USA
[6]University of Texas Austin, Austin, TX, USA
[7]University of Michigan, Ann Arbor, MI, USA
[8]Fraunhofer Institute for Laser Technology, Germany
[9]University of California Los Angeles, Los Angeles, CA, USA
[10]nLight Inc., Vancouver, WA, USA
[11]Raytum Photonics LLC, Sterling, VA, USA
[12]Colorado State University, Fort Collins, CO, USA
[13]University of Alabama at Birmingham, Birmingham, AL, USA
[14]Brookhaven National Laboratory. Upton, NY, USA
[15]Northrop Grumman, Redondo Beach, CA, USA
[16]Few Cycle, Inc., Varennes, QC, Canada
[17]Stony Brook University, Stony Brook, NY, USA

Corresponding author email: kiani2@llnl.gov, tongzhou@lbl.gov

Executive summary

Large-scale laser facilities are needed to advance the energy frontier in high energy physics (HEP) and accelerator physics. Laser plasma accelerators (LPAs) are core to advanced accelerator concepts aimed at reaching TeV electron-electron colliders. In these facilities, intense laser pulses drive plasmas and are used to accelerate electrons to high energies in remarkably short distances. A laser-plasma accelerator could in principle reach high energies with an accelerating length that is ~1000× shorter than in conventional RF based accelerators [1]. Notionally, laser driven particle beam energies could scale beyond today's state of the art conventional accelerators. LPAs have produced multi-GeV electron beams in ~20 cm with relative energy spread of about 2% [2], supported by highly developed laser technology [3]. This validates key elements of the US DOE strategy for such accelerators to enable future colliders [4] but extending today's best results to a TeV collider will require lasers with higher average power. Snowmass whitepapers in the AF6 group are being prepared on advanced accelerator concepts like Laser Wakefield Acceleration (LWFA) lead author: Carl Schroeder, Plasma Wakefield Acceleration (PWFA) lead author: Spencer Gessner and Laser driven structures lead author: Joel England. While the per pulse energies envisioned for laser driven colliders are achievable with today's lasers, low laser repetition rates limit potential collider luminosity. Applications will require rates of kHz to tens of kHz at Joules of energy and high efficiency, and a collider would require ~100 such stages, a leap from current Hz-class LPAs. This represents a challenging 1000-fold increase in laser repetition rates beyond current state of the art. This whitepaper describes current research and outlook for candidate laser systems as well as the accompanying broadband and high damage threshold optics needed for driving future advanced accelerators.

In the near term, diode pumped, actively cooled Ti:sapphire systems [5] offer a path to higher rates. This will enable active correction for instabilities such as pointing, pulse energy/shape, and machine learning [6,6] optimization to realize precision LPA potential. However, to reach collider luminosities, repetition rates would need to increase to tens of kHz and efficiencies to tens of percent which requires new laser technologies. As an example, the required laser parameters, for both 1- and 2-µm laser wavelengths for a 1-TeV center-of-mass linear collider design [8] based on laser-driven hollow-plasma channels operating at a plasma density of $10^{17}$ cm-3, are shown below in Table I. These parameters may be scaled to different operating densities or laser wavelengths [9].

|  | 1 µm lasers | 2 µm lasers |
| --- | --- | --- |
| Energy [J] | 6.5 | 1.6 |
| I [$10^{18}$ W/cm²] | 2 | 0.5 |
| Duration, FWHM [fs] | 130 | 130 |
| Peak Power [TW] | 50 | 12 |
| Repetition Rate [kHz] | 47 | 47 |
| Average Power [kW] | 310 | 75 |
| No. LPA stages | 100 | 400 |
| Wall Plug Efficiency [%] | >30 | >30 |
| Contrast | $10^6$ | $10^5$ |

Table 1: Nominal laser parameters for a near-term 1-GeV, 1-kHz LPA demonstrator facility [11]

High average power ultrafast laser technologies for driving future advanced accelerators

Several technologies have been proposed to achieve these goals, but all need sustained R&D to reach technical readiness as LPA drivers [10, 11, 12]. Further development will also require ongoing research in novel materials, optics, and techniques to support laser systems that exceed present optical damage thresholds. All realistic approaches should be pursued. Many of these new laser designs are highly efficient which is important to make future LPAs economical. They also improve laser spatial and temporal pulse qualities, which in turn enables improvements to the particle beam properties [6, 13].

Intense lasers and compact particle accelerators are also important in national security, medicine and industry, including driving compact secondary particle or photon sources for a variety of applications. Future compact LPAs can enable secondary sources of photons with low energy spread [14], that could deliver high-brightness mono-energetic X-rays useful for detection of nuclear material for instance in cargo inspection [15, 16] and higher resolution medical imaging [17, 18]. Compact accelerators with MeV to GeV laser-driven beams could also potentially replace conventional pulsed-RF driven multi-MeV electron beam accelerators for radiography. Laser driven electron beam and proton beam therapies could be more readily available to clinicians [19, 20]. THz radiation generation from LPAs could be used to improve semiconductor manufacturing [21, 22]. Furthermore, high impact pump-probe experiments for studying ultrafast phenomena are enabled by Joule-class kHz lasers advancing along the LPA development roadmap.

LPA is an essential technical approach for next-generation particle accelerators relevant for the HEP Energy Frontier. Promising laser technologies exist but require additional R&D to realize LPA drivers capable of enabling TeV colliders. Based on science enabled by domestic engineering efforts, laser facilities development in Europe and Asia has put leadership in high-intensity lasers overseas. AS recommended in the 2018 report Opportunities in Intense Ultrafast Lasers: Reaching for the Brightest Light, the United States has an opportunity to make investments in high average power, high intensity lasers to gain leadership in this technological research area [23]. Several workshops on lasers for HEP have considered technology roadmaps for collides and a high flux radiation sources [10,11,24,24,26]. An example prototyping strategy from the 2017 Laser Technology for k-BELLA and Beyond Report is detailed below in Table 2. Central to these roadmaps are near- and mid- term laser prototypes that enable validation of power handling systems, feedback controls and high repetition rate operations/data acquisition for the laser as well as the LPA. These laser prototypes can be considered end-products and serve as long-term sources for a wide variety of science experiments or be designed with future upgrades in mind to address roadmap goals. A facility enabling near term kHz experiments that also couples developing laser technologies with LPA experiments is a clear, important and cost-effective next step that will provide a way to investigate challenges and advance the path towards applications.

|  | Near term | Medium Term | Long term |
|---|---|---|---|
| Energy [J] | 3 | 3 | 3 |
| Duration, FWHM [fs] | 30 | 30-100 | 100 |
| Repetition Rate [kHz] | 1 | 10 | 100 |
| Average Power [kW] | 3 | 30 | 300 |
| Rough Order of Magnitude Cost [$M] | 100 | 500 | >1000 |

Table 2: Example laser technology development goals [11]

High average power ultrafast laser technologies for driving future advanced accelerators

Table of Contents

High average power ultrafast laser technologies for driving future advanced accelerators

1. **Ti:Sapphire systems are ready for precision accelerator, laser and science capabilities needed in the near-term**

Continued progress along the LPA development roadmap requires that we both increase laser and LPA performance towards theoretical limits, which requires active correction, and that we address the challenges of operation at high rates while using those rates to enable early applications such as photon sources [4,27]. A near-term kHz system is needed to address these requirements. Ti:sapphire is the only mature laser technology that can realize the required laser at the 3 Joule, 30 femtosecond level and can be built on the five year time scale [10, 11, 12]. Scaling the LPA driver laser to kHz repetition rate and corresponding average power level relies on four key elements: diode pump lasers, cryogenically cooled Ti:sapphire amplifiers, laser beam/pulse stabilization based on active feedback, and low-loss dielectric pulse compression gratings. This is a major milestone on the US plasma accelerator roadmap [4] and timely execution is critical to continue strong progress on that roadmap. One approach uses OPCPA front end, cryogenically cooled Ti:sapphire, and incoherently combined fiber pump lasers. The hybrid OPCPA/Ti:sapphire design has the potential for scaling by an order of magnitude in power increasing the pump rate to 10 kHz with the thousands of fiber-based pump lasers [11]. At the same time, moving towards even higher repetition rates (10-50kHz) will require new laser technologies due to fundamental limits in Ti:sapphire properties. A facility should support both a near term kHz system and future multi-kHz technologies.

2. **Efficient lasers for driving future LPAs beyond kHz require ongoing development**

Support is needed for focused laser R&D to push beyond current laser driver limitations. Diode-pumped solid-state lasers (DPSSLs) are particularly attractive for high efficiencies and robustness. Multiple DPSSL-based laser technologies have been proposed as mid-term solutions towards higher-efficiency laser drivers that can address HEP community needs in the sub-100fs, few J regime. All candidate techniques require development of active control techniques to realize stable and controllable pulse parameters for efficient high-quality LPAs. Development of high peak and average power handling optics will be required for high rate LPAs as well. Furthermore, the need for broadly available modeling software that captures self consistently the required physics of gain, thermal loading and lensing, spectral shaping and other effects required to quantitatively design such lasers has been highlighted in several recent community planning activities [10, 11, 12].

**Error! Reference source not found.** Beam Combined Fibers

Fiber lasers are the most efficient high power lasers demonstrated to date, due to direct diode pumping and very low quantum defect (particularly for 1-micron operation), as well as the fiber geometry where the optical signal and pump interaction is maximized. The large surface area to volume ratio also facilitates rapid heat dissipation and allows very high average powers. The tradeoff with the fiber geometry sets limits on achieving high pulse energies from chirped-pulse amplification (CPA) of ultrashort pulses in fibers, where nonlinear effects at high peak power result in pulse recompression distortions.

The fiber laser path to achieving the high pulse energies and high average powers needed for driving laser-plasma accelerators (LPA) is through combining many ultrashort pulses generated from fiber amplifier arrays, in space, time, and spectrum [11, 23]. A fiber LPA driver is based on a combination of temporal pulse stacking [28] and spatial beam combining [29] to achieve LPA relevant pulse energies, and spectral combining [30] to achieve the short pulse durations needed by LPAs. Besides offering high wall-plug efficiencies and simple thermal management, fiber lasers also provide high beam quality and compact, robust monolithic laser architectures. This fiber laser combination approach requires additional R&D to demonstrate further power and energy scaling, while maintaining robust and high-fidelity coherently-combined output pulses. The baseline technology operates at 1μm wavelength, and 2μm is an alternate.

Spatial beam combining of multiple ultrashort-pulse laser beams into a single beam, providing simultaneous laser average power and pulse energy scaling, was first demonstrated more than a decade ago. Coherent beam combining of two femtosecond fiber chirped-pulse amplifiers was demonstrated using polarizing and interferometric beam splitters in 2010 and 2011, respectively [31, 32]. Femtosecond pulse beam combining of four parallel chirped-pulse fiber amplifier channels was reported in 2012 [33]. Since then this research topic has been rapidly developed over the past decade or so. Coherent beam combining of 61 femtosecond fiber chirped-pulse amplifiers in a tiled-aperture configuration at ~1kW laser average power was reported in 2020 [34]. An ultrafast fiber laser delivering 10.4 kW average output power based on a coherent combination of 12 step-index fiber amplifiers was demonstrated in 2020 [35]. Meanwhile, over the past years, Lawrence Berkeley National Laboratory has developed a novel diffractive combining scheme capable of combining large numbers of femtosecond pulse beams using only two diffractive optics [29], and demonstrated both 1-D and 2-D diffractive combination of 120fs pulsed beams [29,36].

Temporal pulse stacking using cascaded reflecting optical cavities to transform a sequence of phase- and amplitude-modulated laser pulses into a single output pulse was first demonstrated by University of Michigan in 2015 [2814]. By stacking a burst of many amplified laser pulses from a fiber CPA channel, this scheme largely increases achievable pulse energies from a single ultrafast fiber amplifier, eventually enable the full extraction of stored energy in such a fiber amplifier. In 2017, University of Michigan reported coherent stacking of 81ns effectively-long burst of amplified chirped pulses with a burst energy of 10mJ with low nonlinearity from a Yb-doped 85μm core fiber system [37].

Spectral beam combining of ultrafast fiber lasers synthesizes femtosecond pulse spectra from parallel fiber chirped pulse amplifiers, each amplifying different ultrashort-pulse spectrum. This is to address the limitation associated with gain spectral narrowing when amplifying broadband ultrashort optical pulses. This scheme simultaneously increases ultrashort pulse energy and reduces pulse duration, and it was first demonstrated in 2013 where three parallel fiber chirped pulse amplifiers operating at different spectrum were spectrally combined [30]. In 2015, spectral division and combining with a compact multicore ytterbium doped fiber amplifier demonstrated 100fs pulse duration [38].

Since 2014, supported by Department of Energy (DOE) Office of Science (SC) Accelerator Stewardship program, a collaboration between Lawrence Berkeley National Laboratory (LBNL), the University of Michigan (UM), and Lawrence Livermore National Laboratory (LLNL) has been developing multi-dimensional coherent combining techniques towards high peak and high average power ultrafast lasers aimed at driving LPAs and future laser-driven collider stages. The long term collaboration focuses on the pulse energy and average power scalability of coherent-combined ultrafast fiber amplifier arrays simultaneously implementing spatial beam combining and temporal pulse stacking. Additionally, in 2017 and 2020, LBNL has been awarded two DOE SC Early Career Research projects to specifically develop phase and system control for multi-dimensional laser coherent combining, and to develop spectral beam combining and integrate it with spatial & temporal pulse combining towards LPA relevant pulse durations and energies. Significant progresses haven been made under the umbrella of projects, including: (1) developing novel temporal coherent pulse stacking scheme [39], demonstrating new

High average power ultrafast laser technologies for driving future advanced accelerators stacking control [39, 40], and achieving full extraction of stored energy in a high power fiber amplifier [37]; (2) developing novel spatial diffractive pulse combining scheme [29] and showing 2-D ultrashort pulse combining [36], 81-beam diffractive combining [41], deterministic and machine learning control schemes [42, 43]; (3) achieving 22mJ coherent beam combining from three 85µm core chirally-coupled core fiber amplifiers that allows for temporal pulse stacking [44]. It is worth noting that multi-dimensional coherent combining of ultrafast fiber lasers has also been developed worldwide for broader applications besides laser-driven accelerators [45], e.g. laser machining and processing, to fully leverage the technical advantages of ultrafast fiber lasers.

The fiber approach towards LPA applications requires additional R&D to demonstrate further power and energy scaling, while maintaining robust and high-fidelity coherently combined output, and pulse duration reduction to 10s of fs. To further address this, with newly awarded support from the Gordon and Betty Moore Foundation via Berkeley Lab Foundation from 2021, LBNL will collaborate with UM and LLNL to build and demonstrate a multidimensional coherently-combined fiber laser system that delivers ultrashort pulses at ~200mJ pulse energy, ~30fs pulse duration, and ~1kW average power, and to demonstrate an LPA with this laser system, in a ~5 year time frame. This work will serve as a demonstrator on the energy and power scalability of the coherently-combined fiber laser approach towards facility-scale LPA drivers.

In 5-10 years, the technology is expected to be further developed with more support, scaling up in laser pulse energy and average power (to Joules and ~10 kW) and engineered for more compact, robust, and stable systems (a natural advantage of fiber lasers). This will lead to 10 kW class LPA facilities, e.g. full kBELLA design specs [11], that can deliver GeV level electron beams for high-energy applications, including high brightness radiation sources and LPA staging/cascading experiments. The well-developed technology will also facilitate broader applications including cancer treatment and industrial/security detection.

In the long term (>15 years), this scalable technology is expected to reach multi-J laser energies, 100's kW laser powers, 100 kHz laser and LPA rep-rates, multi-GeV LPA energies, enabling powerful accelerators to study fundamental science through high-energy, high-luminosity experiments. Furthermore, by coupling together (staging) many fiber-driven LPA modules, we expect an LPA-based TeV e-/e+ collider, envisioned by the community and DOE [4], will become a viable approach based on this scalable technology. A very large-scale project like a TeV collider would require high wall-plug efficiencies (tens of percent), and fiber lasers can inherently provide these efficiencies.

Important milestones (for the next >15 years) towards LPA and collider laser drivers include achieving:
  I. >100 mJ energy, 30 fs duration, multi-kHz rep-rate laser pulses to demonstrate a fiber-laser-driven LPA operating in the resonant regime at high plasma density, enabling intermediate light source applications including LPA-based, multi-kHz, high-brightness gamma and x-ray sources;
  II. Sub-Joule energy, 30 fs duration, multi-kHz rep-rate laser pulses to realize a fiber ("mini-kBELLA") LPA facility, delivering hundreds of MeV electron beam energies [46];
  III. Multi-Joule energy, 100 fs duration, multi-kHz rep-rate laser pulses, in a facility-grade system, to demonstrate the pulse energy and power scalability, with the pulse duration needed for LPA staging in an LPA-based lepton collider [8];
  IV. Multi-Joule energy, 30 fs duration, multi-kHz rep-rate laser pulses to realize a fiber ("kBELLA") LPA facility, delivering GeV electron beam energies [11];
  V. Collider stage laser driver: multi-Joule energy, tens of kHz rep-rate (>100 kW average power), 30 fs (for injector LPA) and 100 fs (for staging LPA) durations [4].

High average power ultrafast laser technologies for driving future advanced accelerators

Existing funding (multi-$M level) is expected to allow achieving Milestones A (>100 mJ level) in the next 4-5 years. The resources needed to achieve Milestone B is at the few tens of $M level, with a planned timeline of 6-8 years. Milestones C and D would require funding support at the $100M level and is likely to span into the next 10 years. Additional R&D on studying energy/power scaling and duration reduction challenges and risks will need to be further carried out to realize the milestones, and the investments needed are included in the aforementioned estimated funding support levels.

2.2 Big Aperture Tm:YLF

The Big Aperture Thulium (BAT) laser concept, developed by LLNL, is based on implementing directly diode-pumped Tm:YLF in the gas-cooled slab architecture employed in the High repetition rate Advanced Petawatt Laser System (HAPLS) [47] and Mercury [48] lasers. The radiative lifetime of Tm:YLF is ~15 ms [49], an order of magnitude longer than that of typical Yb-doped materials, and nearly two orders of magnitude greater than that of Nd-doped materials. This long lifetime opens a new regime of operation for high energy lasers, termed multi-pulse extraction (MPE), in which efficient pulsed operation is achieved with continuous pumping while maintain a steady state population inversion. Tm:YLF has strong absorption near $\lambda \approx 790$nm where high power laser diodes are highly developed and commercially available. This broad absorption spectrum combined with a dominant cross-relaxation mechanism that effectively results in two ions excited to the laser upper level per absorbed pump photon supports highly efficient laser operation at a center wavelength near $\lambda \approx 1.9\mu m$. The Tm:YLF stimulated emission spectrum is sufficiently broad to support the amplification of high energy pulses with bandwidth compressible to sub-100fs pulse durations.

Recently, LLNL has demonstrated the generation of 4J, ns-duration pulses and the energy extraction approaching 50 J in with longer duration pulses [50] from a compact, diode-pumped Tm:YLF amplifier Work is ongoing to further extend these results to higher pulse energies and to ultrashort pulse durations. Furthermore, LLNL is currently further scaling the gas-cooling technique for this material system that will enable high average power required for LPA applications, as well as improved understanding of material properties under use conditions. Working in collaboration with OSU, multilayer dielectric gratings (MLD) suitable for pulse compression of high energy, high average power lasers near at wavelength near 1.9µm [51] were developed. These gratings were demonstrated to have femtosecond damage thresholds greater than 100 mJ/cm2 for ~50fs duration pulses.

2.3 Thin-disk lasers for Multi-Pulse LWFA

The energy required to drive a large-amplitude plasma wave can be delivered over many plasma periods, rather than in a single period, if the driving pulse is modulated. The essential idea of the multi-pulse laser wakefield accelerator (MP-LWFA) is to deliver the driving laser energy over several plasma periods. Doing this reduces the demands placed on the laser system, and allows the use of emerging laser technologies able to deliver the required drive energy at multi-kilohertz repetition rates, and with wall-plug efficiencies at least two orders of magnitude higher than the lasers used today. The required modulation can be achieved in a two-step process: (i) spectral modulation of the long drive pulse by co-propagation with a low amplitude plasma wave driven by a short, low-energy seed pulse; (ii) conversion of the spectral modulation to temporal modulation via a dispersive optic to generate a train of short pulses suitable for resonantly driving a plasma accelerator.

This concept is closely related to the beat-wave accelerator investigated in the 1980 and 1990s. A major difference, however, is that that earlier work used nanosecond-duration lasers, corresponding to more than $10^4$ modulations of the driver; this in turn placed very tight tolerances on the plasma density required for resonant excitation. Finding and maintaining resonance was made extremely difficult by the very low (<< 1 Hz) pulse repetition rate of the lasers employed. The use of picosecond-duration, kilohertz repetition rate lasers would make it trivial to tune and lock to resonant excitation.

Thin-disk laser technology which may be suitable for this approach. The optical-to-optical efficiency of thin disk lasers exceeds 50%, and recently thin-disk lasers have produced pulse energies of about 1 J at a repetition rate of 1 kHz [52, 53]. The picosecond duration of the pulses produced by these lasers is too long to drive a laser-plasma accelerator directly, but the pulses can be modulated to form a train of short laser pulses suitable for resonant wakefield excitation [54, 55]. Existing, efficient thin-disk lasers can be used to accelerate electrons to GeV level energies at kHz-repetition-rate.

Researchers at Oxford University recently proposed how picosecond duration laser pulses of this type could be modulated in a three-step process to drive a GeV-scale MP-LWFA [55]. In the first stage, the modulator stage, a leading short ($\tau_{seed} \leq T_{p0}/2$), low-energy seed pulse and a trailing long ($\tau_{drive} \gg T_{p0}$), high-energy driving laser pulse are guided by a plasma channel. Here $T_{p0}$ refers to the plasma period on axis. The seed pulse drives a low amplitude plasma wave which periodically modulates the temporal phase of the drive pulse, and hence generates frequency sidebands at $\omega_0 + m\omega_{p0}$ (where $\omega_0$ is the laser central frequency and $\omega_{p0}$ is the plasma frequency on axis). If isolated, the red- (m<0) and blueshifted (m>0) sidebands would each form a train of pulses of separation $T_{p0}$. However, the red-and blue-shifted pulse trains are shifted with respect to each other by $T_{p0}/2$, so that the temporal profile of the drive pulse remains smooth. In the second stage, after removal of the seed pulse, the spectral modulation of the drive pulse is converted to a temporal modulation by using a dispersive optical system to introduce a shift of $T_{p0}/2$ between the red- and blue-shifted trains [55]. In the third (accelerator) stage, the pulse train is focused into a second plasma waveguide with the same axial density as the modulator, which resonantly excites a large-amplitude plasma wave for acceleration of externally injected electrons. Particle in cell (PIC) simulations of this concept demonstrate that a 1.7 J, 1 ps driver, and a 140 mJ, 40 fs seed pulse can accelerate electrons to an energy of 0.65 GeV in a 100-mm-long plasma channel with axial density of 2.5 x $10^{17}$ cm$^{-3}$ [55].

Note that resonant excitation of a wakefield by a train of ~ 7 laser pulses has been demonstrated experimentally [54], as has a first step towards recovery of unused wakefield energy. Further, in addition to the thin-disk lasers mentioned above, the other key components required to realize this new approach have all recently been demonstrated, and, in principle, all are capable of multi-kilohertz operation. Sub-50 fs, 100 mJ lasers pulses which would be suitable as seed pulses have been demonstrated [56, 57], and meter-scale, low-loss, all optical plasma waveguides [58-64] with on-axis densities of order $10^{17}$ cm$^{-3}$ have been developed. In recent work, one of these hydrodynamic optical-field-ionized (HOFI) plasma channels was successfully operated at a kHz repetition rate for a duration of more than 6 hours [65].

In the next 5 years, it will be critical to achieve an experimental demonstration of all the key stages of our concept for driving an LPA with modulated ps-duration pulses from a thin-disk laser [55]. This work will include demonstration of spectral modulation of a ps-duration driver pulse by a seed-driven plasma wave; demonstration that this modulation can be converted to a temporal modulation by a dispersive optical system; and demonstration that the temporally-modulated pulses can drive a plasma wave. In addition, we will seek to develop the auxiliary components required for an LPA, such as high-repetition-rate (~ kHz) plasma source, diagnostics, and active feedback and stabilization systems. On a 10 years perspective we foresee the need to construct a GeV level demonstrator operating at 1 kHz rep rate. We would expect success in achieving this to lead to many applications with the potential to transfer some of this technology to industry. On the greater than 10 year time-frame we would envisage R&D needed to stage many kHz, GeV-scale modules, as well as work to increase the bunch luminosity. However, accurate predictions for what can be achieved on this time-scale are difficult. A clearer picture will be developed as results are obtained over the next 5 years.

The spectral bandwidth of gain media like Yb:YAG is limiting the generation of short pulses at high peak power. A very efficient way of generating shorter pulses with pulse duration <<100fs starting with the typical output of Yb:YAG or Yb doped fiber lasers of multi 100 fs has been demonstrated in the past years [66]. Efficiencies of about 95% have been demonstrated at kW average output power [67]. Currently scaling of the pulse energy to the single Joule class is investigated. This concept could transfer the high efficiency and average power capability of well established concepts like Fiber, Innoslab and ThinDisk to the required pulse duration and peak power for generating the pulse parameters required for LPA. Currently this concept is also investigated for operation at 2 µm wavelength. On a 5 years perspective we expect to demonstrate the energy scaling to J class at reasonable size of the cell.

Another proposed path to efficient multi-Joule sub-100 fs lasers operating at multi-kHz repetition rate relies on diode pumped cryogenically cooled Yb:YAG as the amplifier gain medium. Cryo-cooled Yb:YAG laser technology demonstrated by Colorado State University (CSU) and XUV Lasers [68] has already been used to generate >1 J pulses at 1 kHz repetition rate that were compressed into pulses of < 5ps duration. However, the gain bandwidth of Yb:YAG limits the pulsewidth of these cryogenically cooled kW average power lasers to > 3 ps. Spectral broadening and compression of Yb:YAG ps pulses in long, large diameter hollow-core fibers can provide a path forward for the direct generation of sub-100 fs high energy laser pulses at high average powers [69, 70]. A collaboration between CSU and Few-Cycle is currently investigating how to combine all performance aspects into one system [71].

## 3. Novel materials, optics, and techniques for high damage threshold

One of the main reasons for pursuing laser-based particle accelerators is its tremendous potential to reduce the size and cost of building and operating accelerators, as lasers can achieve > GeV/cm acceleration fields. To attain the goal of next generation laser-based particle accelerators, one needs to build petawatt class (PW) lasers reliably operating at repetition rates kHz or greater, where the word 'reliably' is the key. Current state of the art petawatt class lasers [72] can semi-reliably operate at single shot mode (see www.lasernetus.org, for example), and a few lasers (Aleph laser ABL and HAPLS) [5, 73] can achieve ultraintense experimental operation rate > 1 Hz for limited amount of time. The main bottleneck in achieving prolonged reliable operation of these class of lasers at high repetition rate is optical component failure in the laser chain, especially during and post-compression stages. Among the optical components, the weakest link has been identified to be pulse compression gratings, and to a lesser extent, high reflector mirrors and antireflection coatings on the amplifier crystals of these chirped pulse amplification (CPA) lasers. Whereas high reflector mirrors can operate at several hundred mJ/cm$^2$ fluence at high repetition rate, for 100 fs pulses, most gratings for PW class systems operate around 100 mJ/cm$^2$ [74]. Ideally, a 1 J/cm$^2$ high repetition rate operational fluence for femtosecond pulses would be a sufficient condition to make these lasers operationally reliable, thereby rendering them to be used in HEP class facilities.

### 3.1 Gratings

Currently for PW class laser systems (pulse duration range of 30-120 fs, and an energy range of 7 – 100 J/pulse), grating sizes are typically 0.5 m or larger in size [72, 74], operating in repetition rates between single shot and 1 Hz, costing > $100,000 and six months to a year to fabricate. The goal would be to create reliable new gratings that would be at least 3x smaller, reducing costs by 5-10x, and production time by 3x. There are three paths available, with the first two being commercially available and deployed in high power laser systems around the world.

1. Gold coated gratings
2. Dielectric gratings
3. Hybrid gratings

Gold coated gratings with Au coatings on patterned photo-resist have been around for many decades, and their single pulse laser induced damage threshold (LIDT) for 100 fs pulses have been reliably measured to be ~0.4 J/cm$^2$ [75]. Conformal Au coated gratings where a conformal coating of Au is applied onto a fused silica etched grating, has been shown to have an LIDT of ~0.25 J/cm$^2$ for as short as 25 fs pulses [76], for which they have the necessary bandwidth to support. They have the advantage of being recoated after damage (underlying silica structure survives a much higher fluence) relatively inexpensively (approximately 10% of the cost of a new grating), but it does not solve the reliability problem. However, there is a theoretical limit to what Au coating can survive in a single shot femtosecond pulse, which is ~0.4 J/cm$^2$. So, Au coating may not be the answer.

Di-electric gratings hold much promise[74, 77], as low index component (e.g. fused silica) materials in the interference coating layers can survive > 1 J/cm$^2$ fluences at even < 10 fs pulse durations [78,79]. However, typically the high index component(s) (lower bandgap, susceptible to much lower LIDT) do(es) not survive such high fluences (LIDT lowers with lowering of bandgap [80]). Moreover, typical di-electric grating designs do not support the large bandwidth required for < 100 fs pulses, and those that do uses different high index layers, some of which lowers the LIDT of such systems. For 500 fs or longer pulse durations, which require small bandwidth, there gratings have been shown to have LIDTs >1 J/cm$^2$ [74]. The other factor is field intensification inside the structure due to standing wave modes inside the multi-layered structures, which are affected simultaneously by ionization due to the strong field of the laser. Until recently, such processes were never modeled dynamically [81, 82], and recent modeling development shows that even though the materials consisting of individual layers may have higher LIDT, the intensification inside gratings allow significant free carrier generation, thereby significantly lowering the LIDT of these grating systems [82]. A sustained modeling and optimization coupled with grating fabrication and LIDT testing for at least a five year period is necessary to explore how far this di-electric approach can be pushed for PW class laser operations.

Hybrid gratings, where a metal layer buried under few layers of dielectric materials, may offer best of both worlds. They offer significantly higher bandwidth and lower (high order) dispersion than pure dielectric gratings, and may exhibit higher LIDT than pure metal (Au) coated gratings [74]. Such gratings would be ideal for < 50 fs operation in PW class lasers. No dynamic modeling incorporating ionization has been developed yet for this class of gratings. Currently, no coordinated efforts exist to model, optimize, fabricate and test these gratings. A longer termed sustained effort would be needed for this grating technology to mature to be used in laser based accelerators.

3.2 Mirrors
Currently, dielectric interference coating (IC) based high reflectors (mirrors) at pulse durations between 40-100 fs ranges can achieve ~1 J/cm2 LIDT [83], and thereby, can operate reliably at around 0.5 J/cm$^2$ for single shot to Hz level operation. The LIDT of mirror coatings is a major constraint for reducing the footprint of mirrors or other components in a CPA laser system. In fact, the limitation of the grating LIDT does not allow reducing the size of the mirrors post-compression. If gratings were made to achieve the LIDT of the mirrors, then the size, cost, complexity and reliability of the PW class lasers would be enhanced tremendously. It is however, not clear how well any of these would survive at kHz or higher repetition rate operation. We discuss these issues in next section.

There are strategies to design high reflectors with large bandwidth to support propagation of femtosecond pulses and at the same time increase LIDT. Among these strategies are the use of two high index layer materials in the mirror coating and thickness control in the multilayer design to shift the peak of the standing wave electric field into the low index material in the stack which has higher LIDT [84]. There are opportunities to improve LIDT by exploring new materials, as for example mixtures of $SiO_2$ or $GeO_2$ with $TiO_2$ to selectively design the refractive index to be used as high index material in a multilayer coating. Stress control in the multilayer coating may also play a role for improving LIDT.

3.3 Antireflection coatings

Antireflection coatings (AR) in femtosecond CPA PW lasers are used in the facets of the crystal amplifiers, in windows, beamsplitters and lenses in the CPA optical systems. To meet the bandwidth requirements to support propagation of femtosecond pulses, AR coatings typically consist of about 10 layers of alternating high and low refractive index materials, in contrast to a few layers as used in nanosecond pulse duration lasers. ARs are a bottleneck in PW systems as they typically damage at lower fluence than mirror coatings. Furthermore, their LIDT is influenced by the type and preparation of the surface of the substrate in which they are deposited [85]. Similarly strategies as in the mirror design which employ several high index materials are at hand. Nonetheless, the interplay of the LIDT of the individual layers in the stack and that of the substrate/coating interface make it difficult to identify design characteristics that will universally apply to AR designs. Significant research is required to identify processes and mechanisms that are affecting the AR coatings LIDT. Similar to mirrors, the LIDT of AR is also affected by the repetition rate at which the laser operates.

High repetition rate operation: It is well-known that di-electric oxide materials used in interference dielectric coatings (ICs) ICs exhibit lowering of LIDT with multiple pulse interaction, sometimes as much as by 50% [86]. It is widely accepted that the effect is due to various defects, color centers[87], voids [88, 89] and trap generation in these amorphous materials from successive pulses that causes this [87]. However, this laser induced effect has not been quantitatively studied even in pure materials, let alone in IC systems. This is because the effect is complex with multiple variables [90–92](e.g. coating methods affecting micro and nano-structuring of materials and interfaces between layers, annealing, stress in coatings, impurities in coatings, packaging, storing of coated systems) affecting the outcome. A systematic study of the LIDT of individual material coatings developed and characterized with tightly controlled and repeatable growth parameters incorporating quantitative detection and characterization of defect generation (using state of the art defect characterization techniques, such as transmission electron microscopy (TEM) and scanning tunneling electron microscopy (STEM), depth resolved cathodoluminescence (DRCLS), etc.) inside these materials followed by establishing the mechanism of lowering LIDT via such defects would be an essential first step. Next step would be to carry this level of detailed analysis on actual IC systems and observe how layers behave under multiple laser irradiation, whether there is defect and trap migration between layers affecting LIDT performance, what is the origin of blister formation [93, 94], which seems to be one of the preferred methods of pre-ablative damage in both dielectric mirrors and gratings. Once the multipulse damage mechanisms are

High average power ultrafast laser technologies for driving future advanced accelerators comprehensively understood, sustained efforts in materials development and coatings will be needed to develop next generation coatings for kHz or higher repetition rate lasers to be used in next generation laser-based accelerators and colliders.

**4. Multi-terawatt long wavelength Infrared lasers for HEP and NP frontiers**

4.1 Laser applications to HEP and NP frontiers

The advent of tabletop laser-driven lepton and proton/ion accelerators in the last three decades ranks among the crowning achievements of chirped-pulse amplified (CPA) laser technology and forms a major component of the scientific impact that earned its inventors the 2018 Nobel Prize in Physics. These accelerators have advanced rapidly since the first ~10 MeV electron "dream beams" were generated in 2004 [95, 96, 97]. Today's CPA-laser-driven wakefield accelerators impart ~10 GeV energy gain to lepton beams within a few centimeters and have become an integral part of federal-agency-level roadmaps for 21st century accelerator science in the U.S. [98], Europe [99] and the U.K. [100]. Nevertheless, other properties of laser-accelerated lepton beams that are critical to applications --- energy spread (10%), charge (tens of pC), spin-polarization (non-existent) --- remain little improved over 2004 capabilities. The small size (tens of microns) of plasma accelerator structures, the resulting high field gradients inside them and the difficulty of controlling lepton injection into them are the main reasons for these shortcomings.

The main motivation for developing high-power mid-wave infrared (MWIR) (3-8 mm) and long-wave infrared (LWIR) (8-15 mm) CPA lasers is to take advantage of favorable wavelength scaling of several physical processes that underlie advanced lepton- and ion- accelerators and radiation sources. In low-density plasmas, ne ~$10^{16}$ cm$^{-3}$, long-wavelength (10 μm) pulses smoothly self-focus to $a_0$ ~ 5 at peak power ~10 TW, and can efficiently drive LWFAs in the blowout regime at a density two orders of magnitude below those typically seen in NIR experiments resulting in plasma bubbles of 1000 times bigger volume. Electrons injected into these supersized bubbles see smaller internal field gradients that otherwise depolarize and energy-broaden accelerated beams. This enables the acceleration of ultra-short spin-polarized electron bunches (e-bunches) with sub-% energy spread, as required for future electron-positron and gamma-gamma ($\gamma\gamma$) colliders, free electron lasers and cargo interrogators [101]. Producing such bubbles with a NIR laser will require the multi-petawatt peak power. Furthermore, simulations show that the LWIR-driven blowout regime can be free from self-injection [102]. The lack of such "dark current" in the LWIR-driven wakefield acceleration opens up the possibility of injecting electrons with high precision and control either internally from the background plasma via optical two-color ionization [103, 104] (See the Novel Particle Sources white paper for more details, lead author: Matthias Fuchs) or externally from a synchronized conventional linac [105], following bunch compression to fs duration.

Laser-driven ion accelerators will also benefit from the favorable long-wavelength scaling of the ponderomotive force and the critical plasma density. This opens the possibility of accelerating monoenergetic high-energy proton beams by collisionless electrostatic shocks in ionized-gaseous plasma that allows for scaling targetry repetition rate and might become a clean source of light ions, e.g., deuterons, alpha-particles, polarized He. The sources serve for radiation cancer therapy. For both ion- and electron-acceleration processes the required LWIR laser must have 10-100 TW peak power in a picosecond (for ion acceleration) or subpicosecond (for LWFA) pulse length and operate at 10-100 Hz repetition rate [10]. At present, short pulse $CO_2$ laser systems with their ability to store a great deal of energy and handle heat balance at 10+ kW average power due to recirculating gas medium seem to be the primary candidate as a laser-driver source.

Another employment for these sources (particularly LWIR) in HEP is the efficient generation of intense gamma rays through inverse Compton backscattering in colliders which are considered valuable extensions to lepton colliders with distinct advantages. For example, the cross sections for production of charged scalar, lepton, and top pairs are higher in collisions by nearly an order of magnitude compared with $e^-/e^+$ collisions. For WW production, this factor could even be up to 20 [106]. Moreover, the control of polarization of the Compton photons will allow for the verification of the circularly polarized nature of Higgs bosons. With the large number of photons produced per unit of laser energy, an LWIR source interacting with an electron beam whose beta function was optimized for the Rayleigh range of the laser can create significantly more Compton photons than an NIR-based source of the same power [107]. Using an LWIR laser also leads to the suppression of pair production during Compton collisions since the applicable criteria $\lambda[\mu m] > 4.2\, E_e[TeV]$ implies that a laser with $\lambda = 9.2\ \mu m$ is needed for an electron beam with $E_e = 2.2\ TeV$. Efficient conversion of an electron beam with 2.2 TeV of energy to a photon beam with $E \approx 1.6\ TeV$ can be realized using a $CO_2$ laser with τ ~ 70 ps and ~27 J per pulse ($a_0^2$=0.1) [108], which is presently obtainable.

4.2 Mid-IR and long wavelength infrared lasers

There are several methods of producing low-energy, ultra-short laser pulses in LWIR including optical slicing from a long-pulse conventional 1-atm $CO_2$ lasers [109, 110] and optical parametric amplifiers (OPA) [111]. However, electrical-discharge $CO_2$ gas laser amplifiers are currently the only means for attaining terawatt-class peak powers in ultrashort LWIR pulses. The broad bandwidth required for picosecond pulse amplification in $CO_2$ lasers is attainable at *multi-atmospheric* working pressures when collisionally broadened discrete rotational lines in the gain spectrum come to overlap with each other. Optimal conditions for achieving a smooth continuous gain spectrum are achieved when isotopically enriched $CO_2$ gas mixture (e.g., with added stable $^{18}O$ isotope) is used. By these approaches, 15 TW peak power in 3 ps pulses was achieved using large regular-$CO_2$ amplifiers [112] and 5 TW, 2 ps pulses were demonstrated in a more compact laser amplifier with mixed-isotope $CO_2$ content [113].

In MWIR spectral domain, laser devices based on solid-state active medium can be used. Transition metal (TM)-doped II-VI chalcogenides, being middle infrared analogs of Ti-sapphire, are the materials of choice for direct lasing and chirped pulse amplification (CPA) in this spectral range [114, 115]. They enable broad spectral tunability, high peak and average power and ultra-short pulse duration. Another promising approach for generating high-average power and high-peak power MWIR pulses is Optical Parametric Chirped Pulse Amplification (OPCPA). Unlike lasers, pump absorption by the active media is not required in OPA, which minimizes the thermal effects and allows high-average power operation. Although LWIR and MWIR sources are at a lower technical readiness level than NIR technologies and substantial technical challenges remain in particular with respect to high-repetition rate reliable operation, significant progress has been made in recent years and the required enabling technologies for these laser sources are actively under investigation.

Development of the LWIR pulses to the readiness level needed for advanced HEP and NP projects and applications consists of two components: First, the present-day single-pulse performance needs to be upgraded to provide sub-ps pulses with tens of TW of power, which is required to reach the laser strength for driving LWFA experiments in the blowout regime [102]. Second, methods will need to be developed to enable the delivery of these pulses at high repetition rates needed to meet requirements for most practical applications. Currently, both thrusts are under active development.

The first thrust is attainable in the next few years. Specifically, simulations show that the 5 TW laser can be straightforwardly upgraded to 500 fs, 20 TW regime using a high-energy (~10 mJ) seed pulse which is under development at the Accelerator Test Facility at Brookhaven National Laboratory [116]. There are two possible paths for this technology: scaling an all-solid-state OPCPA to a few mJ level or using a 15-20 atm compact $CO_2$ amplifier filled with a broadband multi-isotope mixture. To reach sub-ps operation regime, a post-compression technique that leverages the nonlinear interaction of an LWIR pulse with bulk medium is currently being investigated at ATF. Post-compression of a 2 ps pulse to <500 fs in bulk solid-state materials was recently experimentally demonstrated at a sub-terawatt level [117]; possibility of post-compression of a 3.5 ps, 1 TW pulse to <100 fs in a multi-pass gas cell was demonstrated through numerical modeling [118].

The second thrust, i.e. development of high-average power, high repetition rate ultra-fast LWIR sources presents a bigger challenge. This requires significant improvement of existing high-voltage components used in gas-discharge [108]. Direct optical pumping of the $CO_2$ molecules using high-power NIR and MWIR sources is another promising approach. Optical pumping at $\lambda$=4.3 µm, which is preferable for best quantum efficiency, by a Fe:ZnSe laser [119, 120] and at 2.8 µm with Cr,Er:YSGG- or Ho:YAG-pumped OPA [121] are being investigated. Apart from enabling stable high-repetition operation, optical pumping considerably increases our ability to optimize the pressure and the composition of the active gas mixture. For instance, much higher concentrations of $CO_2$ than 2—5 % typical for discharge-pumped lasers will become possible resulting in compact amplifiers comparable in size to those of solid-state lasers.

A new promising way to build MWIR high-power ultra-fast laser systems is based on solid-state laser architecture utilizing Cr doped II-VI crystals with emission bandwidth supporting single cycle pulse durations. Kerr-lens mode-locked polycrystalline Cr:ZnS(Se) seed lasers, developed during the past years are efficient, compact, and reliable, providing, by far, the best combinations of the output laser parameters in the spectral range between 2 and 3 µm and in a broad range of pulse repetition frequencies [114]. Several groups have demonstrated femtosecond $Cr^{2+}$:ZnSe CPAs at kHz repetition rate with pulse energy approaching 10 millijoule-level [105, 115, 122]. The microsecond-level upper state lifetime of Cr:ZnSe indicates that ns lasers for CPA pumping synchronized to the seed pulses are required. Recently proposed and demonstrated directly diode pumped Big Aperture Thulium (BAT) laser concept enabling ns multi-joule-level systems operating near maximum of Cr:ZnSe absorption band at ~ 1.9 µm at average powers approaching 100 kW [50, 123]. Ultrafast Cr:ZnS(Se) seed lasers in combination with BAT pumped Cr:ZnSe CPA spinning ring architecture is a clear route to reach high-average power and high-peak power MWIR pulses. Using this spinning ring platform 0.15 kW Cr:ZnSe output power at 0.2 kW of Tm-fiber pumping was realized. It was also demonstrated that the proposed architecture practically eliminates thermal lensing effects in Cr:ZnSe and enables unprecedented levels of output power with very high (>60%, close to theoretical maximum) optical-to-optical efficiency [114, 124]. Current Cr:ZnSe gain element technology is also ready for power scaling to the multi-kW level.

Another frontier of the TM:II-VI laser technology is the development of ultrafast lasers and amplifiers based on Fe-doped ZnSe operating over 4-5 µm spectral range optically pumped by the 2.7-3 µm sources based on BAT-spinning ring Cr:ZnSe ns tandem. A unique feature of Fe:ZnSe gain medium is its excellent energy storage capacity at cryogenic temperatures (57 µs luminescence lifetime at 77 K [114]). Therefore, cryogenic Fe:ZnSe amplifiers can be optically pumped by readily available, low-cost free-running Er:YAG lasers with J-level, µs-pulses at the 2.94 µm wavelength. The main design consideration in cryogenic Fe:ZnSe amplifiers is the management of amplified spontaneous emission.

With respect to the MWIR OPCPAs, their success requires high-average power pump lasers at long wavelengths. Diode-pumped sub-nanosecond Tm:YLF lasers at 1.9 µm is a potential candidate as the pump laser. The optical parametric crystals with large aperture, high laser-induced damage threshold, and large nonlinearity are needed. $ZnGeP_2$ (ZGP) is a top choice but other emerging crystals such as $BaGa_2GeSe_6$(BGGSe) should also be considered. One of the grand challenges is the fabrication of sufficient large-size ZGP crystals with minimum residual pump absorption.

Provided appropriate funding, achievement of 10—20 TW peak power in 100—500 fs LWIR pulses can be expected within 5 years. A critical part of this development is the deployment of a high-energy (~10 mJ) LWIR solid-state seed OPCPA or Raman converter. Ongoing R&D is aimed to proof-of-principle demonstration of the GW peak power optically pumped $CO_2$ laser amplifier within the next 3 years. Next goal would be the scale-up this approach to the multi-TW level that requires construction of a ~1 kJ MWIR pump laser (or a laser array) that can be completed on 5—10 years' time scale. Development of the first operation-class 100 TW system can be targeted for 2030's; it very likely will be a hybrid system using both discharge pumped and optically pumped technologies.

Achievement of 2 TW peak power in 50-100 fs MWIR pulses of Cr:ZnSe and Fe:ZnSe lasers centered at 2.5 and 4.5 µm at 10 Hz repetition rate can be expected within 5 years. A critical part of this progress is development of the adequate pump sources, optimization of the large-scale gain element growth technology, improvement of their surface optical damage threshold and suppression of amplified spontaneous emission. Development of the 10 TW class solid-state system operating at kHz repetition rate can be targeted for 2030's. A critical part of this progress will be development of adequate BAT pump sources.

In summary, MWIR and LWIR sources provide unique advantages for high energy physics (HEP) and nuclear physics (NP) applications. The wavelength scaling of the underlying physical processes allows infrared lasers to efficiently drive highly nonlinear large plasma waves, which are ideal for accelerating electrons, while preserving their energy spread and polarization. Similarly, high-energy "$\gamma$" rays, such as those needed for a $\gamma\gamma$ collider, can be created through Compton Backscattering of electrons with MWIR/LWIR laser sources. The use of LWIR sources for this application allows significant suppression of pair production, an undesirable byproduct of the laser-electron beam interaction. The interaction of these lasers with an overcritical density plasma (which occurs at gaseous densities for LWIR wavelengths) has been shown to generate tens of MeV ion beams, which are of interest for NP experiments. Terawatt class laser pulses at 2-3 ps have already been demonstrated and an active R&D effort is underway to deliver tens of TW of power within the next 5 years through a combination of energy increase and pulse-length reduction. A critical goal is to deliver reliable, high-repetition rate LWIR and MWIR lasers during the 2030's.

High average power ultrafast laser technologies for driving future advanced accelerators

**References**

1. W. Leemans and E. Esarey. "Laser-driven plasma-wave electron accelerators." Phys. Today 62, no. 3 (2009).

2. A. J. Gonsalves, K. Nakamura, J. Daniels, C. Benedetti, C. Pieronek, T. C. H. De Raadt, S. Steinke et al. "Petawatt laser guiding and electron beam acceleration to 8 GeV in a laser-heated capillary discharge waveguide." Physical review letters 122, no. 8 (2019).

3. K. Nakamura, H.S. Mao, A.J. Gonsalves, H. Vincenti, D.E Mittelberger, J. Daniels, A. Magana, C. Toth, and W.P. Leemans, Diagnostics, control and performance parameters for the BELLA high repetition rate petawatt class laser. *IEEE Journal of Quantum Electronics*, *53*(4), pp.1-21. (2017).

4. "Advanced Accelerator Development Strategy Report: DOE Advanced Accelerator Concepts Research Roadmap Workshop". United States https://www.osti.gov/servlets/purl/1358081. (2016).

5. E. Sistrunk, T. Spinka, A. Bayramian, S. Betts, R. Bopp, S. Buck, K. Charron et al. "All diode-pumped, high-repetition-rate advanced petawatt laser system (HAPLS)." In *CLEO: Science and Innovations*, pp. STh1L-2. Optical Society of America, (2017).

6. Z. H. He, B. Hou, V. Lebailly, J. A. Nees, K. Krushelnick, and A. G. R. Thomas, "Coherent control of plasma dynamics." Nature communications 6, 1-7 (2015).

7. T. C. Galvin, S. Herriot, B. Ng, W. Williams, S. Talathi, T. Spinka, E. Sistrunk, C. Siders, an C. Haefner, "Active learning with deep Bayesian neural network for laser control," Proc. SPIE, 10751, Optics and Photonics for Information Processing XII, 107510N (2018).

8. C. Schroeder et al., "Laser-plasma-based linear collider using hollow plasma channels," in Proc. 2nd European Advanced Accelerator Concepts Workshop, Nucl. Instrum. Meth. A 829 (2016).

9. C. Schroeder et al., "Physics considerations for laser-plasma linear colliders," Phys. Rev. Special Topics: Accel. Beams **13,** 10 (2010).

10. *Workshop on Laser Technology for Accelerators* (2013); https://projects-web.engr.colostate.edu/accelerator/reports/Lasers_for_Accelerators_Report_Final.pdf

11. Report of Workshop on Laser Technology for k-BELLA and Beyond, (May 9-11, 2017); kBELLA workshop report http://www2.lbl.gov/LBL-Programs/atap/Report_Workshop_k-BELLA_laser_tech_final.pdf

12. R. Falcone, F. Albert, F. Beg, S. Glenzer, T. Ditmire, T. Spinka, and J. Zuegel. "Workshop Report: Brightest Light Initiative (March 27-29 2019, OSA Headquarters, Washington, DC)." *arXiv preprint arXiv:2002.09712* (2020).

13. N. Delbos, C. Werle, I. Dornmair, T. Eichner, L. Hübner, S. Jalas, S. W. Jolly et al. "Lux–A laser–plasma driven undulator beamline." *Nuclear Instruments and Methods in Physics Research Section A: Accelerators, Spectrometers, Detectors and Associated Equipment* 909 (2018).

14. F. Albert and A. G. R Thomas, "Applications of laser wakefield accelerator-based light sources." Plasma Phys. Control. Fusion 58, 103001 (2016).

15. H. E Martz, S. M. Glenn, J. A. Smith, C. J. Divin, and S. G. Azevedo. "Poly-versus mono-energetic dual-spectrum non-intrusive inspection of cargo containers." IEEE Transactions on Nuclear Science 64, 7 (2017).

16. C. G. R. Geddes, S. Rykovanov, N. H. Matlis, S. Steinke, J. Vay, E. H. Esarey, B. Ludewigt et al. "Compact quasi-monoenergetic photon sources from laser-plasma accelerators for nuclear detection and characterization." Nuclear Instruments and Methods in Physics Research Section B: Beam Interactions with Materials and Atoms 350 (2015).

17. S. Cipiccia, S. M.,Wiggins, R. P. Shanks, M. R Islam, G. Vieux, R. C., Issac et alA tuneable ultra-compact high-power, ultra-short pulsed, bright gamma-ray source based on bremsstrahlung radiation from laser-plasma accelerated electrons. *Journal of Applied Physics*, **111**, 6 (2012).

18. G. Sarri, G. D. J. Corvan, W. Schumaker, J. M. Cole, Antonino Di Piazza, H. Ahmed, C. Harvey et al. "Ultrahigh brilliance multi-MeV *γ*-ray beams from nonlinear relativistic Thomson scattering." *Phys. Rev. Lett.* **113**, 224801. (2014).

19. L. Labate, D. Palla, D. Panetta, F. Avella, F. Baffigi, F. Brandi, F. Di Martino et al. "Toward an effective use of laser-driven very high energy electrons for radiotherapy: Feasibility assessment of multi-field and intensity modulation irradiation schemes." *Scientific Reports* 10, no. 1 (2020).

20. V. Malka, S. Fritzler, E. Lefebvre, E. d'Humières, R. Ferrand, G. Grillon, C. Albaret et al. "Practicability of proton therapy using compact laser systems." *Medical physics* **31**, 6 (2004).

21. A. Rousse, C. Rischel, S. Fourmaux, I. Uschmann, S. Sebban, G. Grillon, P. Balcou et al. "Non-thermal melting in semiconductors measured at femtosecond resolution." *Nature* **410**, 6824 (2001).

22. T. Jeon and D. Grischkowsky. "Nature of conduction in doped silicon." *Physical Review Letters* **78**, 6 (1997).

23. P. Bucksbaum, "Opportunities in Intense Ultrafast Lasers-Reaching for the Brightest Light". Report No. DOE-NRC-13488. National Academy of Sciences, Engineering., and Medicine, (2018).

24. Building for Discovery: Strategic Plan for U.S. Particle Physics in the Global Context; May 2014. United States (2014).

25. Accelerating Discovery, A. "Strategic Plan for Accelerator R&D in the US." HEPAP Accelerator R&D Subpanel Report, April (2015).

26. Advanced Accelerator Development Strategy Report: DOE Advanced Accelerator Concepts Research Roadmap Workshop. United States (2016).

27. "Decadal Assessment of Plasma Science" National Academies of Sciences, United States (2020).

28. T. Zhou, J. Ruppe, C. Zhu, I. Hu, J. Nees, and A. Galvanauskas, "Coherent pulse stacking amplification using low-finesse Gires-Tournois interferometers," Optics Express 23, 7442-7462 (2015).

29. T. Zhou, T. Sano, and R. Wilcox, "Coherent combination of ultrashort pulse beams using two diffractive optics," Optics Letters 42, 4422-4425 (2017).

30. W. Chang, T. Zhou, L. A. Siiman, and A. Galvanauskas, "Femtosecond pulse spectral synthesis in coherently-spectrally combined multi-channel fiber chirped pulse amplifiers," Opt. Express 21, 3897-3910 (2013).

31. E. Seise, A. Klenke, J. Limpert, and A. Tünnermann, "Coherent addition of fiber-amplified ultrashort laser pulses," Opt. Express 18, 27827-27835 (2010).

32. L. Daniault, M. Hanna, L. Lombard, Y. Zaouter, E. Mottay, D. Goular, P. Bourdon, F. Druon, and P. Georges, "Coherent beam combining of two femtosecond fiber chirped-pulse amplifiers," Opt. Lett. 36, 621-623 (2011).

33. L. A. Siiman, W.Chang, T. Zhou, and A. Galvanauskas, "Coherent femtosecond pulse combining of multiple parallel chirped pulse fiber amplifiers," Opt. Express 20, 18097-18116 (2012).

34. I. Fsaifes, L. Daniault, S. Bellanger, M. Veinhard, J. Bourderionnet, C. Larat, E. Lallier, E. Durand, A. Brignon, and J. Chanteloup, "Coherent beam combining of 61 femtosecond fiber amplifiers," Opt. Express 28, 20152-20161 (2020).

35. M. Müller, C. Aleshire, A. Klenke, E. Haddad, F. Légaré, A. Tünnermann, and J. Limpert, "10.4 kW coherently combined ultrafast fiber laser," Opt. Lett. 45, 3083-3086 (2020).

36. T. Zhou, Q. Du, T. Sano, R.Wilcox, and W. Leemans, "Two-dimensional combination of eight ultrashort pulsed beams using a diffractive optic pair," Opt. Lett. 43, 3269-3272 (2018)

37. H. Pei, J. Ruppe, S. Chen, M. Sheikhsofla, J. Nees, Y. Yang, R. Wilcox, W. Leemans, and A. Galvanauskas, "10mJ Energy Extraction from Yb-doped 85μm core CCC Fiber using Coherent Pulse Stacking Amplification of fs Pulses," in Laser Congress 2017 (ASSL, LAC), OSA Technical Digest (online) (Optica Publishing Group, 2017), paper AW4A.4.

38. Ph. Rigaud, V. Kermene, G. Bouwmans, L. Bigot, A. Desfarges-Berthelemot, and A. Barthélémy, "Spectral division amplification of a 40 nm bandwidth in a multicore Yb doped fiber and femtosecond pulse synthesis with in-fiber delay line," Opt. Express 23, 27448-27456 (2015)

39. Y. Yang, L. Doolittle, A. Galvanauskas, Q. Du, G. Huang, J. Ruppe, T.Zhou, R. Wilcox, and W. Leemans, "Optical phase control of coherent pulse stacking via modulated impulse response," J. Opt. Soc. Am. B 35, 2081-2090 (2018).

40. Y. Xu, R. Wilcox, J. Byrd, L. Doolittle, Q. Du, G. Huang, Y. Yang et al. "FPGA-based optical cavity phase stabilization for coherent pulse stacking." *IEEE Journal of Quantum Electronics* 54, no. 1 (2017): 1-11.

41. Q. Du, D. Wang, T. Zhou, D. Li, and R. Wilcox, "81-beam coherent combination using a programmable array generator," Opt. Express 29, 5407-5418 (2021).

42. Q. Du, T. Zhou, L. R. Doolittle, G. Huang, D. Li, and R. Wilcox, "Deterministic stabilization of eight-way 2D diffractive beam combining using pattern recognition," Opt. Lett. 44, 4554-4557 (2019).

43. D. Wang, Q. Du, T. Zhou, D. Li, and R.Wilcox, "Stabilization of the 81-channel coherent beam combination using machine learning," Opt. Express 29, 5694-5709 (2021).

44. A. Rainville, M. Chen, M. Whittlesey, Q. Du, and A. Galvanauskas, "22mJ Coherent Beam Combining from Three 85µm Core CCC Fiber Amplifiers," in Conference on Lasers and Electro-Optics, OSA Technical Digest (Optica Publishing Group, 2021), paper SW2B.4.

45. H. Stark, J. Buldt, M. Müller, A. Klenke, A. Tünnermann, and J. Limpert, "23 mJ high-power fiber CPA system using electro-optically controlled divided-pulse amplification," Opt. Lett. 44, 5529-5532 (2019).

46. C.G.R. Geddes et al., "Laser technology for Thomson MeV photon sources based on laser-plasma accelerators," Proc. 2014 Advanced Accel. Conf. 2014, AIP Conference Proceedings 1777 (2016), 110002.

47. E. Sistrunk, T. Spinka, A. Bayramian, S. Betts, R. Bopp, S. Buck, K. Charron, J. Cupal, R. Deri, M. Drouin, A. Erlandson, E. S. Fulkerson, J. Horner, J. Horacek, J. Jarboe, K. Kasl, D. Kim, E. Koh, L. Koubikova, R. Lanning, W. Maranville, C. Marshall, D. Mason, J. Menapace, P. Miller, P. Mazurek, A. Naylon, J. Novak, D. Peceli, P. Rosso, K. Schaffers, D. Smith, J. Stanley, R. Steele, S. Telford, J. Thoma, D. VanBlarcom, J. Weiss, P. Wegner, B. Rus, and C. Haefner, "All Diode-Pumped, High-repetition-rate Advanced Petawatt Laser System (HAPLS)," in Conference on Lasers and Electro-Optics, OSA Technical Digest (online) (Optica Publishing Group, 2017), paper STh1L.2.

48. A. Bayramian, P. Armstrong, E. Ault, R. Beach, C. Bibeau, J. Caird, R. Campbell, B. Chai, J. Dawson, C. Ebbers, A. Erlandson, Y. Fei, B. Freitas, R. Kent, Z. Liao, T. Ladran, J. Menapace, B. Molander, S. Payne, N. Peterson, M. Randles, K. Schaffers, S. Sutton, J. Tassano, S. Telford & E. Utterback The Mercury Project: A High Average Power, Gas-Cooled Laser for Inertial Fusion Energy Development, Fusion Science and Technology, 52:3, 383-387 (2007).

49. B. M. Walsh, N. P. Barnes, and B. Di Bartolo, "Branching ratios, cross sections, and radiative lifetimes of rare earth ions in solids: Application to $Tm^{3+}$ and $Ho^{3+}$ ions in $LiYF_4$," J. Appl. Phys. 83, 2772 (1998).

50. I. Tamer, B. A. Reagan, T. Galvin, J. Galbraith, E. Sistrunk, A. Church, G. Huete, H. Neurath, and T. Spinka, "Demonstration of a compact, multi-joule, diode-pumped Tm:YLF laser," Opt. Lett. 46, 5096-5099 (2021).

51. S. Zhang, M. Tripepi, A. AlShafey, N.Talisa, H. T. Nguyen, B.A. Reagan, E. Sistrunk, D.J. Gibson, D.A. Alessi, and E. A. Chowdhury, "Femtosecond damage experiments and modeling of broadband mid-infrared dielectric diffraction gratings," Opt. Express 29, 39983-39999 (2021).

52. C. Herkommer, P. Krötz, R. Jung, S. Klingebiel, C. Wandt, R. Bessing, P. Walch et al., "Ultrafast thin-disk multipass amplifier with 720 mJ operating at kilohertz repetition rate for applications in atmospheric research.", Opt. Express 28, 30164 (2020).

53. Y. Wang, H. Chi, C. Baumgarten, K. Dehne, A. R. Meadows, A. Davenport, G. Murray, B. A. Reagan, C. S. Menoni, and J. J. Rocca, "1.1 J Yb: YAG picosecond laser at 1 kHz repetition rate." Opt. Lett. 45, 6615 (2020).

54. J. Cowley, C. Thornton, C. Arran, R. J. Shalloo, L. Corner, G. Cheung, C. D. Gregory et al., "Excitation and control of plasma wakefields by multiple laser pulses." Phys. Rev. Lett. 119, 044802, (2017).

55. O. Jakobsson, S.M. Hooker and R. Walczak, Jakobsson, O., S. M. Hooker, and R. Walczak, "GeV-scale accelerators driven by plasma-modulated pulses from kilohertz lasers." Phys. Rev. Lett. 127, 184801 (2021).

56. Thales Laser Group, ALPHA kHz—High Repetition Rate Ti:Sa Laser Series, https://www.thalesgroup.com/en/laserdatasheets.

57. M. Kaumanns, D. Kormin, T. Nubbemeyer, V. Pervak, and S. Karsch, "Spectral broadening of 112 mJ, 1.3 ps pulses at 5 kHz in a LG 10 multipass cell with compressibility to 37 fs." Opt. Lett. **46**, 929, (2021).

58. R. J. Shalloo, C. Arran, L. Corner, J. Holloway, J. Jonnerby, R. Walczak, H. M. Milchberg, and S. M. Hooker, "Hydrodynamic optical-field-ionized plasma channels.", Phys. Rev. E 97, 053203 (2018).

59. R. J. Shalloo, C. Arran, A. Picksley, A. Von Boetticher, L. Corner, J. Holloway, G. Hine et al., "Low-density hydrodynamic optical-field-ionized plasma channels generated with an axicon lens.", Phys. Rev. Accel. Beams **22**, 041302 (2019).

60. S. Smartsev, C. Caizergues, K. Oubrerie, J. Gautier, J. Goddet, A. Tafzi, K. T. Phuoc, V. Malka, and C. Thaury, "Axiparabola: a long-focal-depth, high-resolution mirror for broadband high-intensity lasers.", Opt. Lett. 44, 3414 (2019).

61. A. Picksley, A. Alejo, J. Cowley, N. Bourgeois, L. Corner, L. Feder, J. Holloway et al., "Guiding of high-intensity laser pulses in 100-mm-long hydrodynamic optical-field-ionized plasma channels.", Phys. Rev. Accel. Beams 23, 081303 (2020).

62. A. Picksley, A. Alejo, R. J. Shalloo, C. Arran, A. von Boetticher, L. Corner, J. A. Holloway et al., "Meter-scale conditioned hydrodynamic optical-field-ionized plasma channels.", Phys. Rev. E **102**, 053201 (2020).

63. L. Feder, B. Miao, J. E. Shrock, A. Goffin, and H. M. Milchberg, "Self-waveguiding of relativistic laser pulses in neutral gas channels.", Phys. Rev. Research **2**, 043173 (2020).

64. B. Miao, L. Feder, J. E. Shrock, A. Goffin, and H. M. Milchberg, "Optical guiding in meter-scale plasma waveguides." Phys. Rev. Lett. 125, 074801 (2020).

65. A. Alejo, J. Cowley, A. Picksley, R. Walczak, and S. M. Hooker. "Demonstration of kilohertz operation of hydrodynamic optical-field-ionized plasma channels." *Physical Review Accelerators and Beams* 25, no. 1 (2022).

66. P. Russbueldt, J. Weitenberg, J. Schulte, R. Meyer, C. Meinhardt, H. D. Hoff-mann, and R. Poprawe, "Scalable 30 fs laser source with 530 W average power, " Opt. Lett., 44, 5222-5225 (2019).

67. C. Grebing, M. Müller, J. Buldt, H. Stark, J. Limpert, "Kilowatt-average-power compression of millijoule pulses in a gas-filled multi-pass cell, " Opt. Lett., 45 (22), 6250-6253 (2020).

68. C. Baumgarten, M. Pedicone, H. Bravo, H.C. Wang, L. Yin, C.S. Menoni, J. J. Rocca, and B.A. Reagan, "1 J, 0.5 kHz Repetition Rate Picosecond Laser," Optics Letters, Vol. 41, 3339-3342, (2016).

69. G. Fan, P.A. Carpeggiani, Z. Tao, E. Kaksis, T. Balciunas, G. Coccia, V. Cardin, F. Légaré, B. E. Schmidt, and A. Baltuška, "Post-Compression of pulses from an Yb Amplifier to 25fs at 40mJ," in 2019 Conference on Lasers and Electro-Optics Europe and European Quantum Electronics Conf., OSA Technical Digest (Optical Society of America), (2019).

70. Y.G. Jeong, R. Piccoli, D. Ferachou, V. Cardin, M. Chini, S. Hädrich, J. Limpert, R. Morandotti, F. Legare', B.E. Schmidt, S.L.Razzati; "Direct compression of 170-fs 50-cycle pulses down to 1.5 cycles with 70% transmission" Scientific reports 8, 1, (2018).

71. H. Seiler, S. Palato, B. E. Schmidt, and P. Kambhampati, "Single fiber based solution for coherent multidimensional spectroscopy in the visible regime', Optics Letters., 42, 643, (2017).

72. C. N. Danson, C. Haefner, J. Bromage, T. Butcher, J.-C. F. Chanteloup, E. A. Chowdhury, A. Galvanauskas, L. A. Gizzi, J. Hein, D. I. Hillier, N. W. Hopps, Y. Kato, E. A. Khazanov, R. Kodama, G. Korn, R. Li, Y. Li, J. Limpert, J. Ma, C. H. Nam, D. Neely, D. Papadopoulos, R. R. Penman, L. Qian, J. J. Rocca, A. A. Shaykin, C. W. Siders, C. Spindloe, S. Szatmári, R. M. G. M. Trines, J. Zhu, P. Zhu, and J. D. Zuegel, "Petawatt and exawatt class lasers worldwide," High Power Laser Sci. Eng. (2019).

73. A. Bayramian, R. Bopp, M. Borden, B. Deri, R. DesJardin, J. M. Di Nicola, M. Drouin, A. Erlandson, S. Fulkerson, J. Jarboe, G. Johnson, H. Zhang, B. Heidl, J. Horner, K. Kasl, D. Kim, E. Koh, J. Lusk, C. Marshall, D. Mason, T. Mazanec, J. Naylon, J. Nissen, K. Primdahl, B. Rus, M. Scanlan, K. Schaffers, T. Simon, T. Spinka, J. Stanley, C. Stolz, S. Telford, and C. Haefner, "High energy, high average power, DPSSL system for next generation petawatt laser systems," in 2016 Conference on Lasers and Electro-Optics (CLEO) (2016), pp. 1–2.

74. N. Bonod and J. Neauport, "Diffraction gratings: from principles to applications in high-intensity lasers," Adv. Opt. Photon. 8(1), 156–199 (2016).

75. B. C. Stuart, M. D. Feit, S. Herman, a. M. Rubenchik, B. W. Shore, and M. D. Perry, "Optical ablation by high-power short-pulse lasers," J. Opt. Soc. Am. B 13(2), 459 (1996).

76. P. Poole, S. Trendafilov, G. Shvets, D. Smith, and E. Chowdhury, "Femtosecond laser damage threshold of pulse compression gratings for petawatt scale laser systems.," Opt. Express 21(22), 26341–51 (2013).

77. J. Neauport, E. Lavastre, G. Razé, G. Dupuy, N. Bonod, M. Balas, G. de Villele, J. Flamand, S. Kaladgew, and F. Desserouer, "Effect of electric field on laser induced damage threshold of multilayer dielectric gratings," Opt. Express 15(19), 12508 (2007).

78. B. Chimier, O. Utéza, N. Sanner, M. Sentis, T. Itina, P. Lassonde, F. Légaré, F. Vidal, and J. C. Kieffer, "Damage and ablation thresholds of fused-silica in femtosecond regime," Phys. Rev. B - Condens. Matter Mater. Phys. 84(9), 1–10 (2011).

79. N. Sanner, O. Ut́za, B. Chimier, M. Sentis, P. Lassonde, F. Ĺgaŕ, and J. C. Kieffer, "Toward determinism in surface damaging of dielectrics using few-cycle laser pulses," Appl. Phys. Lett. 96(7), 3–6 (2010).

80. M. Mero, J. Liu, W. Rudolph, D. Ristau, and K. Starke, "Scaling laws of femtosecond laser pulse induced breakdown in oxide films," Phys. Rev. B 71, 1–7 (2005).

81. S. Zhang, A. Davenport, N. Talisa, C. Menoni, V. Gruzdev, and E. Chowdhury, "2D dynamic ionization simulation from ultrashort pulses in multilayer dielectric interference coatings," in Laser-Induced Damage in Optical Materials 2020, C. W. Carr, V. E. Gruzdev, D. Ristau, and C. S. Menoni, eds. (SPIE, 2020).

82. S. Zhang, M. Tripepi, A. AlShafey, N. Talisa, H. Nguyen, B. Reagan, E. F. Link, D. Gibson, D. Alessi, and E. Chowdhury, "Femtosecond damage experiments and modeling of broadband mid-infrared dielectric diffraction gratings," Opt. Express 29(24), 39983–39999 (2021).

83. R. A. Negres, C. J. Stolz, K. R. P. Kafka, E. A. Chowdhury, M. Kirchner, K. Shea, and M. Daly, "40-Fs Broadband Low Dispersion Mirror Thin Film Damage Competition," in Proceedings of SPIE 10014, Laser-Induced Damage in Optical Materials 2016 (2016), 10014, p. 100140E.

84. D. Schiltz, D. Patel, C. Baumgarten, B.A. Reagan, J.J. Rocca, C.S. Menoni, "Strategies to increase laser damage performance of Ta2O5/SiO2 mirrors by modifications of the top layer design," Appl. Opt. 56, C136-C139 (2017).

85. Hanchen Wang, Alexander R. Meadows, Elzbieta Jankowska, Emmett Randel, Brendan A. Reagan, Jorge J. Rocca, and Carmen S. Menoni, "Laser induced damage in coatings for cryogenic Yb:YAG active mirror amplifiers," Opt. Lett. 45, 4476-4479 (2020).

86. M. Mero, B. Clapp, J. C. Jasapara, W. . Rudolph, D. Ristau, K. Starke, J. Krüger, S. Martin, and W. Kautek, "On the damage behavior of dielectric films when illuminated with multiple femtosecond laser pulses," Opt. Eng. 44(5), 051107 (2005).

87. A. Rosenfeld, M. Lorenz, R. Stoian, and D. Ashkenasi, "Ultrashort-laser-pulse damage threshold of transparent materials and the role of incubation," Appl. Phys. A 69(1), S373--S376 (n.d.)

88. C. Wu and L. V. Zhigilei, "Microscopic mechanisms of laser spallation and ablation of metal targets from large-scale molecular dynamics simulations," Appl. Phys. A Mater. Sci. Process. 114(1), 11–32 (2014).

89. C. B. Schaffer, A. O. Jamison, and E. Mazur, "Morphology of femtosecond laser-induced structural changes in bulk transparent materials," Appl. Phys. Lett. 84(9), 1441–1443 (2004).

90. C. J. Stolz, J. E. Wolfe, J. J. Adams, M. G. Menor, N. E. Teslich, P. B. Mirkarimi, J. A. Folta, R. Soufli, C. S. Menoni, and D. Patel, "High laser-resistant multilayer mirrors by nodular defect planarization" Appl. Opt. 53(4), A291--A296 (2014).

91. P. F. Langston, E. Krous, D. Schiltz, D. Patel, L. Emmert, A. Markosyan, B. Reagan, K. Wernsing, Y. Xu, Z. Sun, R. Route, M. M. Fejer, J. J. Rocca, W. Rudolph, and C. S. Menoni, "Point defects in $Sc_2O_3$ thin films by ion beam sputtering," Appl. Opt. 53(4), A276--A280 (2014).

92. E. Jankowska, S. Drobczynski, and C. S. Menoni, "Analysis of surface deformation in thin-film coatings by carrier frequency interferometry," Appl. Opt. 56(4), C60--C64 (2017).

93. N. Talisa, M. Tripepi, B. Harris, A. AlShafey, A. Davenport, E. Randel, C. S. Menoni, and E. Chowdhury, "Single-shot few-cycle pulse laser-induced damage and ablation of $HfO_2$/$SiO_2$-based optical thin films," in Optics InfoBase Conference Papers (2019), Part F129-.

94. N. Talisa, A. Alshafey, M. Tripepi, J. Krebs, A. Davenport, E. Randel, C. S. Menoni, and E. A. Chowdhury, "Comparison of damage and ablation dynamics of thin film interference coatings initiated by few cycle pulses vs longer femtosecond pulses," Opt. Lett. 45(9), 2672 (2020).

95. C. Geddes, C. Toth, J. van Tilborg, J. *et al.* High-quality electron beams from a laser wakefield accelerator using plasma-channel guiding. *Nature* **431,** 538–541 (2004). https://doi.org/10.1038/nature02900

96. J. Faure, Y. Glinec, A. Pukhov, *et al.* A laser–plasma accelerator producing monoenergetic electron beams. *Nature* **431,** 541–544 (2004). https://doi.org/10.1038/nature02963

97. S. Mangles, C. Murphy, Z. Najmudin. *et al.* Monoenergetic beams of relativistic electrons from intense laser–plasma interactions. *Nature* **431,** 535–538 (2004). https://doi.org/10.1038/nature02939

98. U. S. Dept. of Energy, Office of Science, Washington, DC. Advanced Accelerator Development Strategy Report: DOE Advanced Accelerator Concepts Research Roadmap Workshop. https://doi.org/10.2172/1358081 (2016).

99. Cros, B. and Muggli, P. Towards a Proposal for an Advanced Linear Collider: Report on the Advanced and Novel Accelerators for High Energy Physics Roadmap Workshop, CERN, Geneva. web site (2017).

100. Plasma wakefield accelerator research 2019-2040: a community-driven UK roadmap compiled by the Plasma Wakefield Accelerator Steering Committee. http://pwasc.org.uk/uk-roadmap-development (2019).

101. Z. Nie, F. Li, F. Morales, S. Patchkovskii, O. Smirnova, W. An, N. Nambu, D. Matteo, K.A. Marsh, F. Tsung, W.B. Mori, C. Joshi, "In Situ generation of high-energy spin-polarized electrons in a beam-driven plasma wakefield accelerator", Physical Review Letters, 126(054801):1 – 6

102. P. Kumar, K. Yu, R. Zgadzaj, M. Downer, I. Petrushina, R. Samulyak, V. Litvinenko, and N. Vafaei-Najafabadi, "Evolution of the self-injection process in the transition of an LWFA from self-modulation to blowout regime", Physics of Plasmas, 28, 013102 (2021); https://doi.org/10.1063/5.0027167

High average power ultrafast laser technologies for driving future advanced accelerators

103. X. L. Xu, Y. P. Wu, C. J. Zang, F. Li, Y. Wan, J. F. Jua, "Low-emittance electron beam generation frmo a laser wakefield accelerator using two laser pulses with different wavelengths," Phys. Rev. ST Accel. Beams 17, 061301 (2014).

104. L-L. Yu et al., "Two-color Laser-ionization Injection." Physical Review Letters 112, 125001 (2014).

105. Y. Wu, J. Hua, Z.Zhou, J. Zhang, S. Liu, B. Peng, Y. Fang et al. "High-throughput injection–acceleration of electron bunches from a linear accelerator to a laser wakefield accelerator." Nature Physics 17, no. 7 (2021).

106. V. Telnov, arXiv e-prints , hep-ex/9810019 (1998), arXiv:hep-ex/9810019 [hep-ex].

107. I. V. Pogorelsky, M. Polyanskiy, and T. Shaftan, Converting conventional electron accelerators to high peak brilliance Compton light sources, Phys. Rev. Accel. Beams 23, 120702 (2020)

108. I. V. Pogorelsky, M. N. Polyanskiy, and W. D. Kimura, Mid-infrared lasers for energy frontier plasma accelerators, Phys. Rev. Accel. Beams 19, 091001 (2016)

109. A. J. Alcock and P. B. Corkum, Ultra-fast switching of infrared radiation by laser-produced carriers in semiconductors, Can. J. Phys. 57, 1280 (1979).

110. C. V. Filip, R. Narang, S. Y. Tochitsky, C. E. Clayton, and C. Joshi, Optical Kerr switching technique for the production of a picosecond, multiwavelength CO2 laser pulse, Appl. Opt. 41, 3743 (2002).

111. M. N. Polyanskiy and M. Babzien, "Ultrashort Pulses," in CO2 Laser - Optimization and Application, D. C. Dumitras ed. InTech, 139 (2012)

112. D. Haberberger, S. Tochitsky, C. Joshi, "Fifteen terawatt picosecond CO2 laser system," Optics Express 18, 17865 (2010). https://doi.org/10.1364/OE.18.017865

113. M. N. Polyanskiy, I. V. Pogorelsky, M. Babzien, M. A. Palmer, "Demonstration of a 2 ps, 5 TW peak power, long-wave infrared laser based on chirped-pulse amplification with mixed-isotope CO2 amplifiers," OSA Continuum 3, 459 (2020). https://doi.org/10.1364/OSAC.381467

114. Sergey Mirov, Igor Moskalev, Sergey Vasilyev, Viktor Smolski, Vladimir Fedorov, Dmitry Martyshkin, Jeremy Peppers, Mike Mirov, Alex Dergachev, Valentin Gapontsev, "Frontiers of mid-IR lasers based on transition metal doped chalcogenides," IEEE Journal of Selected Topics in Quantum Electronics, 24(5), 1601829, pp 1-29 (2018).

115. Y. Wu, EW Larsen, F. Zhou, L. Wang, A. Marra, D. Schade, J. Li , Z. Chang, "Transition metal doped zinc selenide infrared lasers for ultrafast and intense field science." Emerging Laser Technologies for High-Power and Ultrafast Science (2021).

116. M. N. Polyanskiy, I. V. Pogorelsky, M. Babzien, R. Kupfer, N. Vafaei-Najafabadi, M. A. Palmer, "High-peak-power long-wave infrared lasers with CO2 amplifiers," Photonics 8, 101 (2021). https://doi.org/10.3390/photonics8040101

117. M. N. Polyanskiy, I. V. Pogorelsky, M. Babzien, R. Kupfer, K. L. Vodopyanov, M. A. Palmer, "Post-compression of long-wave infrared 2 picosecond sub-terawatt pulses in bulk materials," Opt. Express 20, 31714 (2021). https://doi.org/10.1364/OE.434238

118. M. G. Hastings, P. Panagiotopoulos, M. Kolesik, V. Hasson, S. Tochitsky, J. V. Moloney, "Few-cycle 10 μm multi-terawatt pulse self-compression in a gas-filled multi-pass cell: a numerical experiment," J. Opt. Soc. Am. B, 39, 266 (2022). https://doi.org/10.1364/JOSAB.437870

119. D. Tovey, S. Y. Tochitsky, J. Pigeon, G. Louwrens, M. Polyanskiy, I. Ben-Zvi, and C. Joshi, "Multi-atmosphere picosecond CO2 amplifier optically pumped at 4.3 μm," Appl. Opt. 58, 5756 (2019). https://doi.org/10.1364/AO.58.005756

120. D. Tovey, J. Pigeon, S. Tochitsky, G. Louwrens, I. Ben-Zvi, D. Martyshkin, V. Fedorov, K. Karki, S. Mirov, and C. Joshi, "Lasing in 15 atm CO2 cell optically pumped by a Fe: ZnSe laser," Opt. Express 29, 31455 (2021). https://doi.org/10.1364/OE.434670

121. V. M. Gordienko , V. T. Platonenko, "Regenerative amplification of picosecond 10-μm pulses in a high-pressure optically pumped CO2 laser," Quantum Electron. 40, 1118 (2010). https://doi.org/10.1070/QE2010v040n12ABEH014426

122. V. E. Leshchenko, B. K. Talbert, Y. H. Lai, S. Li, Y. Tang, S. J. Hageman, G. Smith, P. Agostini, L. F. DiMauro, and C. I. Blaga, "High-power few-cycle Cr: ZnSe mid-infrared source for attosecond soft x-ray physics," Optica 7, 981-988 (2020).

123. C. W. Siders, T. Galvin, A. Erlandson, A. Bayramian, B. Reagan, E. Sistrunk, T. Spinka, and C. Haefner, "Wavelength scaling of laser wakefield acceleration for the EuPRAXIA design point," Instruments 3, 44 (2019).

124. I. Moskalev, S. Mirov, M. Mirov, S. Vasilyev, V. Smolski, A. Zakrevskiy, and V. Gapontsev, "140 W Cr: ZnSe laser system," Optics express 24, 21090-21104 (2016).